\documentclass[11pt]{article}
\usepackage[numbers,sort&compress]{natbib}	
\usepackage{hypernat}				
\usepackage{graphicx}		
\usepackage{amssymb}
\usepackage{simplemargins}
\setallmargins{1in}
\usepackage[american]{babel}
\usepackage{amsmath}
\usepackage{amsfonts}	
\usepackage{amsmath}
\usepackage{url}
\usepackage{multicol}
\usepackage[pdftex]{hyperref}
\hypersetup{
  unicode=false,
  pdffitwindow=true,
  pdfpagelayout=TwoPageLeft,
  pdfhighlight=/P,
  pdfmenubar=false,
  pdftoolbar=false,
  pdfstartview=FitV,
  pdfnewwindow=true,
  colorlinks=true,
  linkcolor=blue,
  anchorcolor=black,
  citecolor=blue,
  filecolor=black,
  menucolor=black,
  urlcolor=blue,
  breaklinks=true,
  pdftitle={On Oscillations in the Social Force Model},
  pdfkeywords={Pedestrian, Crowd, Mass, Simulation, Dynamics, Viswalk, Social Force Model}
  pdfsubject={Social Force Model, Pedestrian, Crowd, Mass, Simulation, Dynamics, Viswalk, Vissim},
  pdfauthor={Kretz},
  pdfcreator={Tobias Kretz},
  pdfproducer={T. Kretz},
}

\title{The Inflection Point of the Speed-Density Relation and the Social Force Model}

\author{Tobias Kretz, Jochen Lohmiller, Johannes Schlaich\\
PTV Group, Haid-und-Neu-Stra{\ss}e 15, D-76131 Karlsruhe, Germany\\
\texttt{\{First.Family\}@ptvgroup.com}\\
}%

\begin{document}
\maketitle

\abstract{
It has been argued that the speed-density digram of pedestrian movement has an inflection point. This inflection point was found empirically in investigations of closed-loop single-file pedestrian movement. The reduced complexity of single-file movement does not only allow a higher precision for the evaluation of empirical data, but it occasionally also allows analytical considerations for micosimulation models. In this way it will be shown that certain (common) variants of the Social Force Model (SFM) do not produce an inflection point in the speed-density diagram if infinitely many pedestrians contribute to the force computed for one pedestrian. We propose a modified Social Force Model that produces the inflection point. 
}

\section{Introduction: Empirical Data on Pedestrians' Speed-Density Relation} \label{sec:1}
\begin{figure}[ht!]
\centering
\includegraphics[height=4cm]{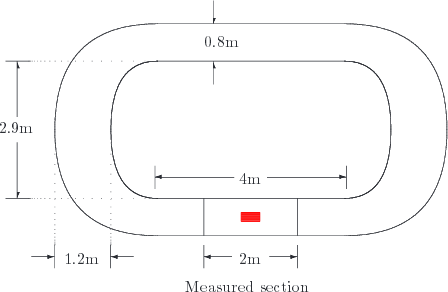} \hspace{12pt}
\includegraphics[height=4cm]{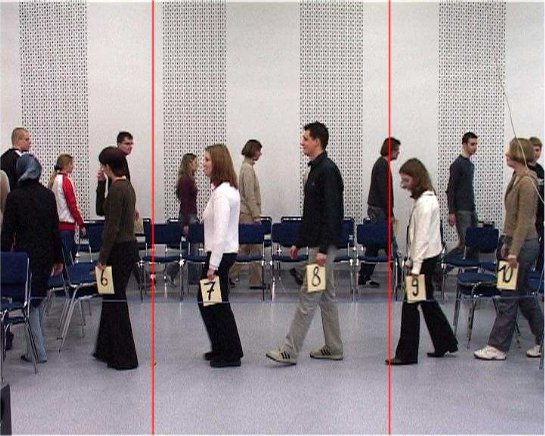}
\caption{Experimental setup. Source: Figures 2 and 3 of \cite{seyfried2005fundamental}}
\label{fig:ExperimentalSetup}       
\end{figure}

In the course of recent years a number of experiments have been conducted in which pedestrians walk single-file in a closed loop \cite{seyfried2005fundamental,chattaraj2009comparison,seyfried2010phase,seyfried2010enhanced,Portz2011analyzing}. Having a different number of pedestrians in the loop, different densities are prepared. In a section of the loop line density and speed are measured. Figure \ref{fig:ExperimentalSetup} shows the experimental setup from which most data stems. There were experiment runs in which the loop was larger and more pedestrians participated, but the principle was always the same.

\begin{figure}[ht!]
\centering
\includegraphics[height=4cm]{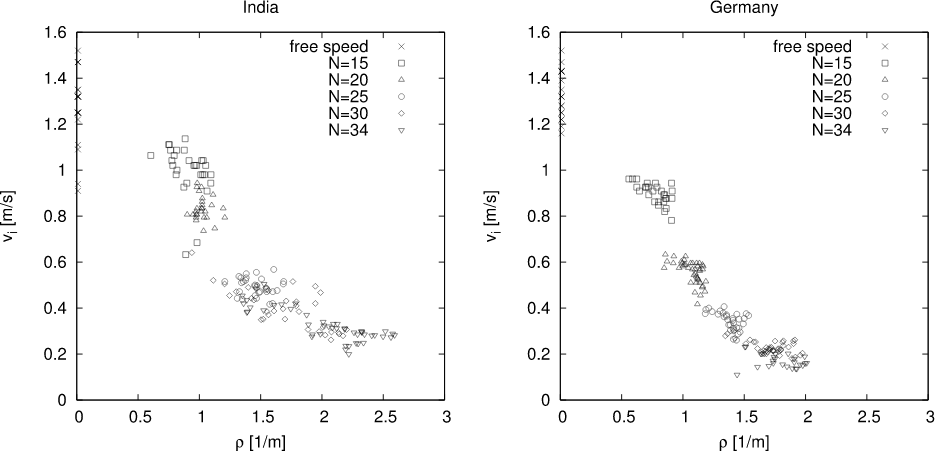}
\caption{Speed density diagram for India (left) and Germany (right). Source: Figure 5 of \cite{chattaraj2009comparison} }
\label{fig:GermanyIndia}       
\end{figure}

The experiment has been conducted at various places around the world. Figure \ref{fig:GermanyIndia} shows the results for India and Germany. With the free speeds plotted in this diagram the existence of an inflection point is obvious. It is not quite clear, however, at which density curvature is maximally negative and also the density of the inflection point can only be estimated roughly. 

Whereas Figure \ref{fig:GermanyIndia} clearly shows that differences between India and Germany were found, the loop size (corridor length) and the profession of participants did not make a difference, at least not an obvious one. Compare Figure 7 of \cite{chattaraj2009comparison} with Figure 1 of \cite{Portz2011analyzing}.

The discussion\footnote{Quote (p. 3): {\em ``Domain I: ... At low densities there is a small and increasing decline of the velocity ... Domain III: ... For growing density the velocity remains nearly constant.''}} in \cite{seyfried2005fundamental} indicates that the existence of an inflection point is common sense and more so that for -- at least moderately -- high densities the curvature of the flow density relation is positive. 

\section{The Social Force Model for Steady-States in Single-File Movement} \label{sec:2}
The circular specification of the Social Force Model \cite{Helbing2000simulating} is defined as\footnote{We neglect here forces from walls from the beginning since we are anyway working towards the one-dimensional case.}:
\begin{eqnarray}
\ddot{\vec{x}}_\alpha(t)=\frac{\vec{v}_{0\alpha}-\dot{\vec{x}}_\alpha(t)}{\tau_\alpha}+\tilde{A}_\alpha \sum_{\beta} w(\vec{x}_\alpha(t),\vec{x}_\beta(t),\dot{\vec{x}}_\alpha(t),\lambda_\alpha) e^{-\frac{|\vec{x}_\beta(t)-\vec{x}_\alpha(t)|-R_\alpha-R_\beta}{B_\alpha}}\hat{e}_{\alpha\beta} \label{eq:circularfull}\\
w(\vec{x}_\alpha(t),\vec{x}_\beta(t),\dot{\vec{x}}_\alpha(t),\lambda_\alpha)=\lambda_\alpha + (1-\lambda_\alpha) \frac{1+\cos(\theta_{\alpha\beta}(\vec{x}_\alpha(t),\vec{x}_\beta(t),\dot{\vec{x}}_\alpha(t)))}{2}
\end{eqnarray}
where $v_{0\alpha}$ is the desired speed of pedestrian $\alpha$. $\tilde{A}_\alpha>0$, $B_\alpha>0$, $0\leq\lambda_\alpha\leq 1$, and $\tau_\alpha>0$ are parameters of the model. $R$ denotes the body radius of a pedestrian. $\hat{e}_{\alpha\beta}$ points from pedestrian $\beta$ on pedestrian $\alpha$. $\vec{x}$ is the position of a pedestrian and dots mark time derivatives (i.e. speed and acceleration). The sum runs over all - potentially infinitely many -- pedestrians in a simulation scenario. Function $w()$ is there to suppress forces acting from behind. In it $\theta_{\alpha\beta}$ is the angle between pedestrian $\alpha$'s velocity vector and the line connecting $\alpha$ and $\beta$.

From here on we assume that parameters $\tilde{A}$, $B$, $\lambda$, $\tau$ $R$, and $v_0$ have identical value for all pedestrians, so we omit the indices. This allows to combine $\tilde{A}$ and $R$ into a new parameter $A=\tilde{A}e^{-2R}$.

Since it is obvious which properties are time dependent, we also omit the ``$(t)$''. Then equation (\ref{eq:circularfull}) reads for the one-dimensional case:
\begin{eqnarray}
\ddot{x}_\alpha             &=& \frac{v_0-\dot{x}_\alpha}{\tau}+A \sum_{\beta} w(x_\alpha,x_\beta,\lambda) e^{-\frac{d_{\alpha\beta}}{B}} \label{eq:circular1d}\\
d_{\alpha\beta}             &=& |x_\beta-x_\alpha| \\
w(x_\alpha,x_\beta,\lambda) &=& \lambda \text{ if } x_\beta-x_\alpha<0\\
w(x_\alpha,x_\beta,\lambda) &=& -1 \text{ if } x_\beta-x_\alpha>0
\end{eqnarray}
with the additional assumption that for all pedestrians and times $\dot{x}>0$. Thus we do not intend to model the loop from the experiment, but assume closed boundary conditions in 1d.

Elliptical specification II is a variant of the Social Force Model where the force between pedestrians -- in addition to the mutual distance -- also depends on the relative velocity of pedestrians $\alpha$ and $\beta$. The full definition is give in \cite{johansson2007specification} and reduced to one dimension it can be found in \cite{Kretz2015oscillations}. If -- as in the scenario discussed in this work -- pedestrians $\alpha$ and $\beta$ have the same velocity -- vanishing relative velocity -- elliptical specification II gives the same force for $\alpha$ as the circular specification. Therefore the further reasoning applies also to elliptical specification II (not elliptical specification I, though).

Now we investigate the steady state of this -- in fact both - model(s). Steady-state means that the speeds and distances remain constant and consequently acceleration is zero for all pedestrians. With the left side of equation (\ref{eq:circular1d}) being zero we can easily compute the steady state speed without having to solve a differential equation:
\begin{equation}
\dot{x}_\alpha = v_0 + \tau A \sum_{\beta} w(x_\alpha,x_\beta,\lambda) e^{-\frac{d_{\alpha\beta}}{B}} \label{eq:steadystate}\\
\end{equation}

As we chose that the parameters for all pedestrians are equal and the system has periodic boundary conditions or is infinitely large, all distances between neighboring pedestrians are equal. Then necessarily the distance (center point to center point) from pedestrian $\alpha$ to the $n$ next neighbor $\beta_n$ can then be written as:
\begin{equation}
d_{\alpha\beta n} = n d_0
\end{equation}

If we resolve the $w()$ function into both directions -- $w()=-1$ for all pedestrians ahead and $w()=\lambda$ for all pedestrians behind -- we can rewrite equation (\ref{eq:steadystate}) more explicitly with the sum running over natural numbers not pedestrians:
\begin{equation}
\dot{x}_\alpha = v_0 - (1-\lambda) \tau A \sum_{n=1}^{\infty}  e^{-\frac{n d_0}{B}}\\
\end{equation}

Since $d_0$ and $B$ both necessarily are positive it is $e^{-\frac{d_0}{B}}<1$ and the sum is the geometric series with the well known result
\begin{eqnarray}
\dot{x}_\alpha &=& v_0 - (1-\lambda) \tau A \left(\frac{1}{1- e^{-\frac{d_0}{B}}}-1\right)\\
               &=& v_0 - (1-\lambda) \tau A \frac{1}{e^{\frac{d_0}{B}}-1}\\
               &=& v_0 - (1-\lambda) \tau A \frac{1}{e^{\frac{1}{B\rho}}-1} \label{eq:steadystatedensity}
\end{eqnarray}
where $\rho$ is the line density of pedestrians $\rho=1/d_0$.

With appropriately chosen values for  parameters $v_0$, $\lambda$, $\tau$ $A$, $B$ equation (\ref{eq:steadystatedensity}) should  be able to reproduce the empirical fundamental diagram as shown in Figure \ref{fig:GermanyIndia} -- obviously not each single data point, but the general, average course. This includes that the speed density relation in equation (\ref{eq:steadystatedensity}) computed for/from the Social Force Model should also yield an inflection point. So for (\ref{eq:steadystatedensity}) we compute the second  derivative of the speed function with regard to density:
\begin{eqnarray}
v(\rho) &=& v_0 - (1-\lambda) \tau A \frac{1}{e^{\frac{1}{B\rho}}-1}\\
\frac{\partial v(\rho)}{\partial \rho} &=& - (1-\lambda) \tau A \frac{e^{\frac{1}{B\rho}}}{B\rho^2(e^{\frac{1}{B\rho}}-1)^2}\\
\frac{\partial^2 v(\rho)}{\partial \rho^2} &=& (1-\lambda) \tau A e^{\frac{1}{B\rho}} \frac{(2B\rho-1)e^{\frac{1}{B\rho}} - (2B\rho+1)}{B^2\rho^4(e^{\frac{1}{B\rho}}-1)^3}
\end{eqnarray}
and test if it is zero for some density $\rho_i$:
\begin{equation}
(2B\rho_i-1)e^{\frac{1}{B\rho_i}} - (2B\rho_i+1) = 0
\end{equation}
This does not have a solution, but the left side of the equation is always negative and approaches zero only asymptotically for $\rho_i \rightarrow \infty$.

At this point we could write down conclusions and title the paper ``Requiem for the Social Force Model'' (and -- btw. -- a number of other models as well). However, experience teaches to first search for possibilities of resurrection. Still we note as first remarkable result of this work: {\em Result 1: Neither the circular specification nor the elliptical specification II of the Social Force Model as originally defined produce an inflection point in the speed density relation for homogeneous steady-state one-dimensional movement.}

The next step does not follow as a consequence of what is written here so far, but it has to be justified a posteriori: we reconsider equation (\ref{eq:steadystate}) and investigate a variant of the model where not all pedestrians, but only the nearest neighbors -- the one in front and the one at the rear -- exert a force on pedestrian $\alpha$. This leads to
\begin{equation}
\dot{x}_\alpha = v_0 - (1-\lambda) \tau A e^{-\frac{1}{B\rho}} \label{eq:steadystatedensityNN}
\end{equation}
which we write seemingly unnecessarily complicated as
\begin{equation}
\dot{x}_\alpha = v_0 - (1-\lambda) \tau A \frac{1}{e^{\frac{1}{B\rho}}-0} \label{eq:steadystatedensityNN2}
\end{equation}
In this form we note that the only difference between equation (\ref{eq:steadystatedensity}) and equation(\ref{eq:steadystatedensityNN2}) -- so between the original formulation where infinitely many pedestrians exert a force on $\alpha$ and the nearest neighbor variant -- is only that in the first case there is a ``1'' and in the second case a ``0'' in the denominator.

Having written the equations in this way the natural next question is ``What if instead of the zero or one in equations (\ref{eq:steadystatedensity}) and (\ref{eq:steadystatedensityNN2}) we write there some $0<k<1$?'':
\begin{equation}
\dot{x}_{k \alpha} = v_0 - (1-\lambda) \tau A \frac{1}{e^{\frac{1}{B\rho}}-k} \label{eq:steadystatedensityk}
\end{equation}
respectively in dimensionless form:
\begin{equation}
f(x)_k(\rho) = 1 - \frac{e^{a}-k}{e^{\frac{a}{x}}-k} \label{eq:steadystatedensityk-ggu}
\end{equation}

This modifies the second derivative:
\begin{eqnarray}
v_k(\rho) &=& v_0 - (1-\lambda) \tau A \frac{1}{e^{\frac{1}{B\rho}}-k} \label{eq:steadystatek}\\
\frac{\partial v_k(\rho)}{\partial \rho} &=& - (1-\lambda) \tau A \frac{e^{\frac{1}{B\rho}}}{B\rho^2(e^{\frac{1}{B\rho}}-k)^2}\\
\frac{\partial^2 v_k(\rho)}{\partial \rho^2} &=& (1-\lambda) \tau A e^{\frac{1}{B\rho}} \frac{(2B\rho-1)e^{\frac{1}{B\rho}} - k(2B\rho+1)}{B^2\rho^4(e^{\frac{1}{B\rho}}-k)^3}
\end{eqnarray}

For $0\leq k < 1$ the new equation to solve is
\begin{equation}
(2B\rho_i-1)e^{\frac{1}{B\rho_i}} - k (2B\rho_i+1) = 0
\end{equation}
and it has a solution. For $k=0$ it is obviously $\rho_i=1/(2B)$. For other values of $k$ the solution needs to be computed numerically. Table \ref{tab:inflectionsolutions} gives some values for $B\rho_i$.

\begin{table}[htbp]
\centering
\caption{Numerical solutions for the value $\rho_i$ of the inflection point with regard to various values for parameter $k$. Values computed with \cite{wolframalpha}}
\begin{tabular}{ll|ll|ll}
\hline\noalign{\smallskip}
$k$&$B\rho_i$&$k$& $B\rho_i$ & $k$ & $B\rho_i$ \\
\noalign{\smallskip}\hline\noalign{\smallskip}
0.0 & 0.500 & 0.90  & 0.981 & 0.99  & 2.049 \\
0.1 & 0.515 & 0.91  & 1.013 & 0.999 & 4.379 \\
0.2 & 0.531 & 0.92  & 1.051 & 0.9999 & 9.416 \\
0.3 & 0.551 & 0.93  & 1.096 & 0.99999 & 20.28\\
0.4 & 0.576 & 0.94  & 1.151 & 0.999999 & 43.68 \\
0.5 & 0.606 & 0.95  & 1.219 & 0.9999999 & 94.10\\
0.6 & 0.646 & 0.96  & 1.309 & 0.99999999 & 202.7\\
0.7 & 0.703 & 0.97  & 1.435 & 0.999999999 & 436.8\\
0.8 & 0.793 & 0.98  & 1.635 & 0.9999999999 & 941.0\\
\noalign{\smallskip}\hline \label{tab:inflectionsolutions}
\end{tabular}
\end{table}

\begin{figure}[ht!]
\centering
\includegraphics[height=3cm]{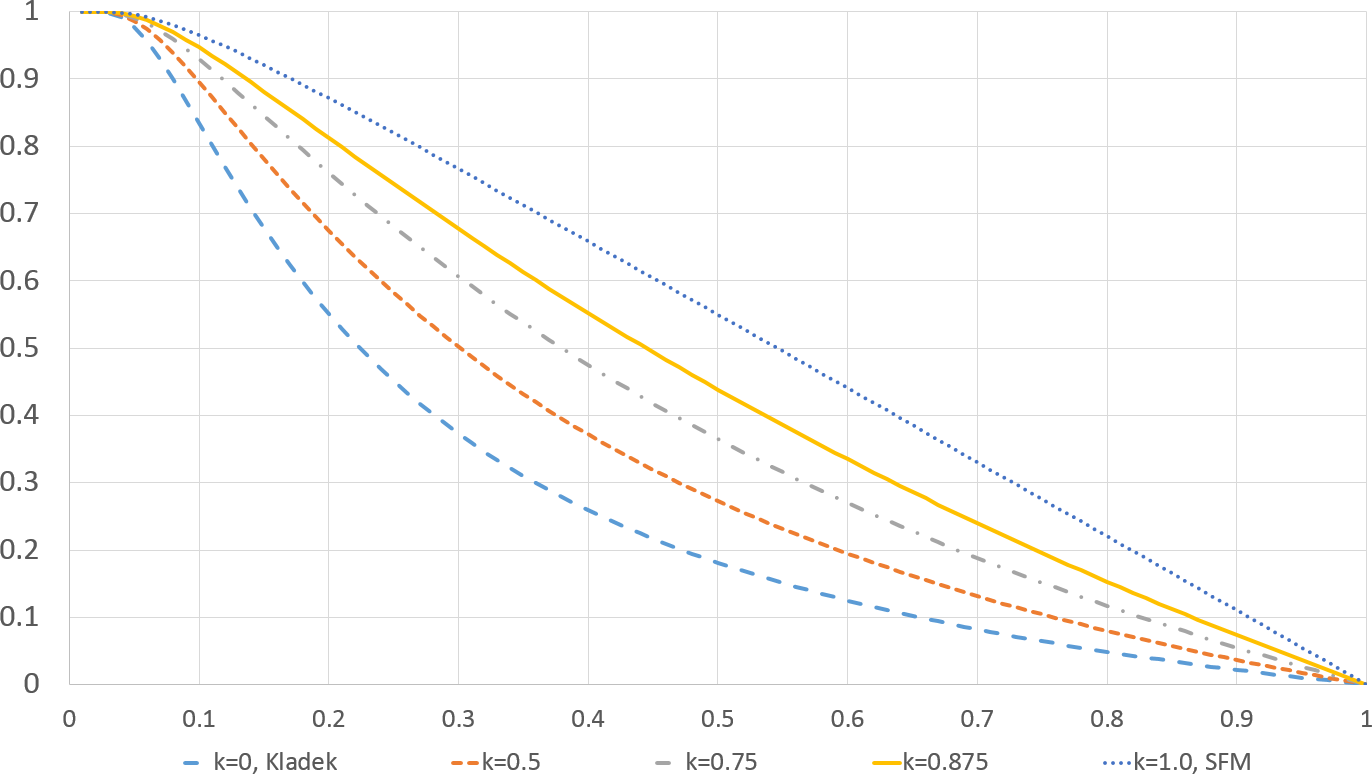} \hspace{12pt}
\includegraphics[height=3cm]{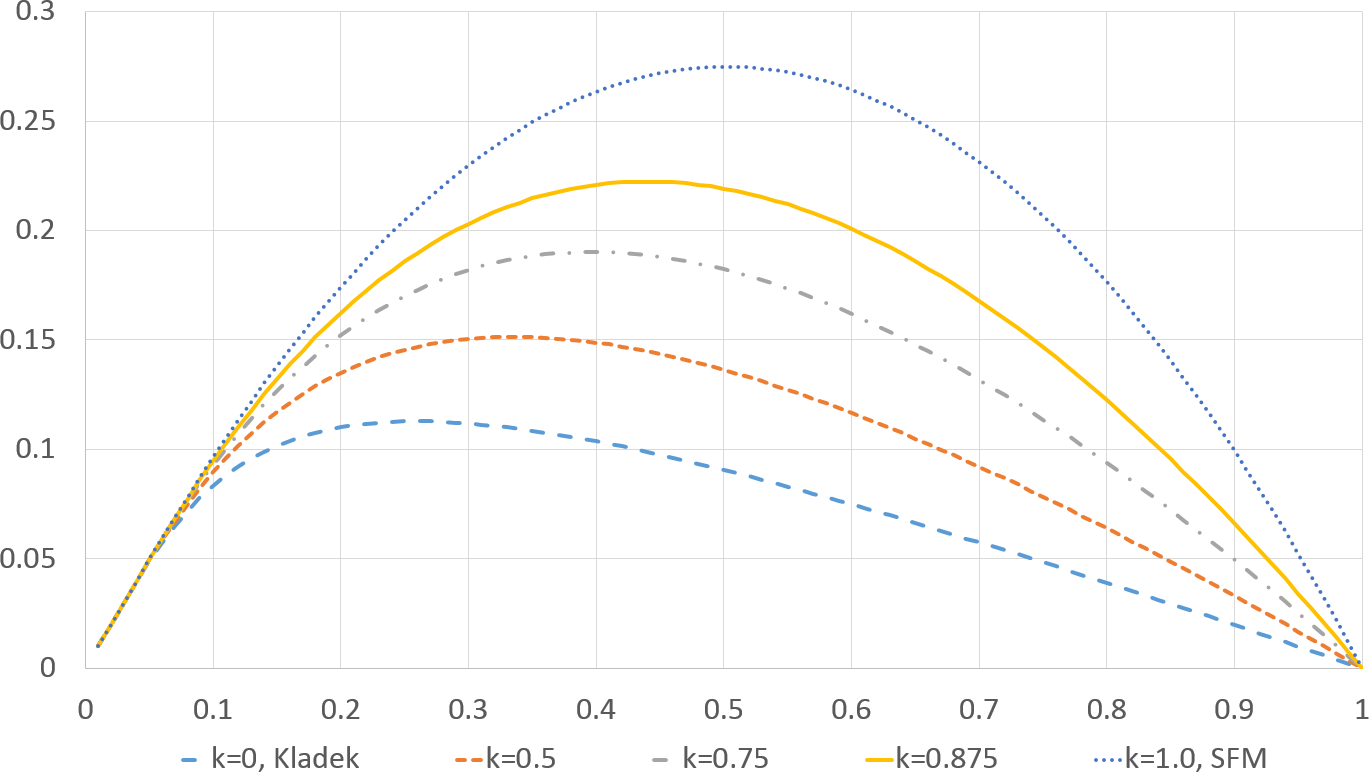}
\caption{Speed-density (left) and flow-density (right) relations of the $k$-extended Social Force Model for various values of parameter $k$. The plots are based on the dimensionless variants of equations and it is $a=0.2$.}
\label{fig:SFMk}       
\end{figure}

The introduction of parameter $k$ produces the desired inflection point in the fundamental diagram. See figure \ref{fig:SFMk}. To find an intuitively comprehensible interpretation for this extension, we write equation (\ref{eq:steadystatek}) in a slightly different manner
\begin{equation}
v_k(\rho) = v_0 - (1-\lambda) \tau A \frac{1}{k}\frac{1}{\frac{e^{\frac{1}{B\rho}}}{k}-1}
\end{equation}
and undo the summation of the geometric series with $k$ ``on the back'' of the exponential function
\begin{eqnarray}
v_k(\rho) &=& v_0 - (1-\lambda) \tau A \frac{1}{k} \sum_{n=1}^{\infty} k^n     e^{-\frac{n}{B\rho}}\\
          &=& v_0 - (1-\lambda) \tau A             \sum_{n=1}^{\infty} k^{n-1} e^{-\frac{n}{B\rho}}
\end{eqnarray}

In terms of forces/acceleration this means:
\begin{equation}
\ddot{x} = \frac{v_0 - v}{\tau} - (1-\lambda) A \sum_{n=1}^{\infty} k^{n-1} e^{-\frac{d_{\alpha\beta}}{B}}
\end{equation}
This is easy to interpret: the next neighbors of $\alpha$ exert a force on $\alpha$ unmodified compared to the original SFM. The second next neighbors exert a force which is suppressed by a factor $k$ (always compared to the original model without parameter $k$). The force from the next to next to next nearest neighbors is suppressed by a factor of $k^2$ and so on. This implies for example that if we take out each second pedestrian from a simulation and the next to next nearest neighbor becomes the next nearest neighbor the force from this new nearest neighbor is larger than before when s/he was just the next to next nearest neighbor although the distance to $\alpha$ is the same as before.

This can be comprehended intuitively. If two pedestrians approach ``me'' independently and the first remote pedestrian overtakes and becomes the closest one ``my'' {\em awareness} is shifted from the former to the new nearest neighbor and that would have an impact on my velocity changes, i.e. acceleration, i.e. forces. This intuitive comprehensibility of the model extension adds to the pleasure of having gained the desired inflection point. Thus:

{\em Result 2: Suppressing the force from each pedestrian in the sequence ordered by distance from the pedestrian for whom forces are calculated with an additional factor $0<k<1$ produces an inflection point in the resulting macroscopic speed-density relation and as a model extension can be motivated intuitively.}

We would like to emphasize that the introduction of parameter $k$ as extension of the Social Force Model brings a major conceptual change. Without it forces of various pedestrians superpose without interfering. With parameter $k$ on the contrary one has to know the local distribution of all pedestrians before one can compute the force of one pedestrian on another one. Forces do not superpose anymore. Instead the extended model -- let us briefly call it ``SFMk'' -- is rather described by Sherif's famous description of social systems: ``…the properties of any part are determined by its membership in the total functional system.'' \cite{sherif1936psychology}. In this sense the SFMk structurally bears some similarities to another extension of the Social Force Model -- namely the dynamic potential \cite{Kretz2011e,Kretz2012e,Kretz2013d,Kretz2014e} -- where the desired direction of pedestrians is computed such that they walk into the direction of earliest expected arrival, for this considering the distribution and movement state of all pedestrians in a holistic way. A difference, however, between SFMk and the dynamic potential is that in SFMk it is exactly known which pedestrian $\beta_i$ contributes what to the effect on the movement of pedestrian $\alpha$. This is not the case for the dynamic potential. We would therefore call the dynamic potential a mesoscopic or mean-field or holistic modeling element while the SFMk is non-superposing but non-holistic and entirely microscopic.

\section{Summary} \label{sec:3}
In this contribution we have computed the macroscopic limit of the Social Force Model of pedestrian dynamics for single-file uni-directional movement. We found that the speed-density curve does not have an inflection point, but that with the introduction of one more parameter to the Social Force Model one can recover an inflection point. The new model intuitively makes sense considering that humans focus their awareness.
%
\bibliographystyle{utphys2011}
\bibliography{author}

\providecommand{\href}[2]{#2}\begingroup\raggedright\begin{thebibliography}{10}

\bibitem{seyfried2005fundamental}
A.~Seyfried, B.~Steffen, W.~Klingsch, and M.~Boltes, ``The fundamental diagram
  of pedestrian movement revisited'',
  \href{http://dx.doi.org/10.1088/1742-5468/2005/10/P10002}{{\em Journal of
  Statistical Mechanics: Theory and Experiment} {\bfseries 2005} no.~10, (2005)
  P10002}, \href{http://arxiv.org/abs/physics/0506170}{{\ttfamily
  arXiv:physics/0506170 [physics]}}.

\bibitem{chattaraj2009comparison}
U.~Chattaraj, A.~Seyfried, and P.~Chakroborty, ``Comparison of pedestrian
  fundamental diagram across cultures'',
  \href{http://dx.doi.org/10.1142/S0219525909002209}{{\em Advances in Complex
  Systems} {\bfseries 12} no.~03, (2009) 393--405},
  \href{http://arxiv.org/abs/0903.0149}{{\ttfamily arXiv:0903.0149
  [physics.soc-ph]}}.

\bibitem{seyfried2010phase}
A.~Seyfried, A.~Portz, and A.~Schadschneider,
  \href{http://dx.doi.org/10.1007/978-3-642-15979-4_53}{``Phase coexistence in
  congested states of pedestrian dynamics'',} in {\em Cellular Automata},
  S.~Bandini, S.~Manzoni, H.~Umeo, and G.~Vizzari, eds., vol.~6350 of {\em
  Lecture Notes in Computer Science}, pp.~496--505.
\newblock Springer Berlin Heidelberg, 2010.
\newblock \href{http://arxiv.org/abs/1006.3546}{{\ttfamily arXiv:1006.3546
  [physics.soc-ph]}}.

\bibitem{seyfried2010enhanced}
A.~Seyfried, M.~Boltes, J.~K{\"a}hler, W.~Klingsch, A.~Portz, T.~Rupprecht,
  A.~Schadschneider, B.~Steffen, and A.~Winkens,
  \href{http://dx.doi.org/10.1007/978-3-642-04504-2_11}{``Enhanced empirical
  data for the fundamental diagram and the flow through bottlenecks'',} in {\em
  Pedestrian and Evacuation Dynamics 2008}, pp.~145--156.
\newblock Springer, 2010.
\newblock \href{http://arxiv.org/abs/0810.1945}{{\ttfamily arXiv:0810.1945
  [physics.soc-ph]}}.

\bibitem{Portz2011analyzing}
A.~Portz and A.~Seyfried,
  \href{http://dx.doi.org/10.1007/978-1-4419-9725-8_52}{``Analyzing Stop-and-Go
  Waves by Experiment and Modeling'',} in {\em Pedestrian and Evacuation
  Dynamics}, R.~D. Peacock, E.~D. Kuligowski, and J.~D. Averill, eds.,
  pp.~577--586.
\newblock Springer US, 2011.
\newblock \href{http://arxiv.org/abs/1003.5446}{{\ttfamily arXiv:1003.5446
  [physics.soc-ph]}}.

\bibitem{Helbing2000simulating}
D.~Helbing, I.~Farkas, and T.~Vicsek, ``{Simulating dynamical features of
  escape panic}'', \href{http://dx.doi.org/10.1038/35035023}{{\em Nature}
  {\bfseries 407} (2000) 487--490},
  \href{http://arxiv.org/abs/cond-mat/0009448}{{\ttfamily
  arXiv:cond-mat/0009448 [cond-mat]}}.

\bibitem{johansson2007specification}
A.~Johansson, D.~Helbing, and P.~Shukla, ``Specification of the social force
  pedestrian model by evolutionary adjustment to video tracking data'',
  \href{http://dx.doi.org/10.1142/S0219525907001355}{{\em Advances in Complex
  Systems} {\bfseries 10} no.~supp02, (2007) 271--288},
  \href{http://arxiv.org/abs/0810.4587}{{\ttfamily arXiv:0810.4587
  [physics.soc-ph]}}.

\bibitem{Kretz2015oscillations}
T.~Kretz, ``On Oscillations in the Social Force Model'', {\em Physica A:
  Statistical Mechanics and its Applications} (2015) ,
  \href{http://arxiv.org/abs/1507.02566}{{\ttfamily arXiv:1507.02566
  [physics.soc-ph]}}.

\bibitem{wolframalpha}
Wolfram|Alpha, 2015.
\newblock Publisher: Wolfram Alpha LLC. Retrieved March 24th 2015.
  \url{https://www.wolframalpha.com/input/?i=solve+%282x-1%29*exp%281%2Fx%29+-+0.9*%282x%2B1%29+for+x}.

\bibitem{sherif1936psychology}
M.~Sherif, {\em The psychology of social norms}.
\newblock Harper, 1936.

\bibitem{Kretz2011e}
T.~Kretz, A.~Gro{\ss}e, S.~Hengst, L.~Kautzsch, A.~Pohlmann, and P.~Vortisch,
  ``{Quickest Paths in Simulations of Pedestrians}'',
  \href{http://dx.doi.org/10.1142/S0219525911003281}{{\em {Advances in Complex
  Systems}} {\bfseries 14} (2011) 733--759},
  \href{http://arxiv.org/abs/1107.2004}{{\ttfamily arXiv:1107.2004
  [physics.soc-ph]}}.

\bibitem{Kretz2012e}
T.~Kretz and A.~Gro{\ss}e, ``{From Unbalanced Initial Occupant Distribution to
  Balanced Exit Usage in a Simulation Model of Pedestrian Dynamics}'', in {\em
  Human Behaviour in Fire Symposium}, {T.J. Shields et al.}, ed., pp.~536--540.
\newblock Interscience Communications, 2012.
\newblock \href{http://arxiv.org/abs/1210.4759}{{\ttfamily arXiv:1210.4759
  [physics.soc-ph]}}.

\bibitem{Kretz2013d}
T.~Kretz, ``{Multi-Directional Flow as Touch-Stone to Assess Models of
  Pedestrian Dynamics}'', in {\em Annual Meeting of the TransportationResearch
  Board 2013}.
\newblock 2013.
\newblock \href{http://arxiv.org/abs/1212.6855}{{\ttfamily arXiv:1212.6855
  [physics.soc-ph]}}.
\newblock on CD: 13-1160. See associated video: http://youtu.be/Ivbstw8FIuo.

\bibitem{Kretz2014e}
T.~Kretz, \href{http://dx.doi.org/{10.1007/978-3-319-02447-9_84}}{``{The Effect
  of Integrating Travel Time}'',} in {\em Pedestrian and Evacuation Dynamics
  2012}, U.~Weidmann, U.~Kirsch, and M.~Schreckenberg, eds., pp.~1013--1027.
\newblock 2013.
\newblock \href{http://arxiv.org/abs/1204.5100}{{\ttfamily arXiv:1204.5100
  [physics.soc-ph]}}.

\end{thebibliography}\endgroup


\end{document}